\documentstyle[twoside,epsfig]{article}                                                                    
\def\3{\ss}                                                                                        
\newcommand{\lmean}{\raisebox{.3ex}{\mbox{$\scriptstyle < $}}}
\newcommand{\rmean}{\raisebox{.3ex}{\mbox{$\scriptstyle > $}}}
\newcommand{\mean}[1]{\lmean\mbox{$ #1 $}\rmean}
\def\Journal#1#2#3#4{{#1} {\bf #2} (#4) #3}
\def\JP{{ J. Phys.\ }}

\def\NIM{ Nucl.\ Instr.\ Meth.\ }
\def\NP{{ Nucl.\ Phys.\ }}
\def\PL{{ Phys.\ Lett.\ }}
\def\PRL{ Phys.\ Rev.\ Lett.\ }
\def\PR{{ Phys.\ Rev.\ }}
\def\ZP{{ Z. Phys.\ }}
\newlength{\gap}
\newcommand{\eT}{\mbox{$E_T$}}
\newcommand{\eTj}{\mbox{$E_T^{\,jet}$}}
\newcommand{\eTg}{\mbox{$E_T^{\,\gamma}$}}

\newcommand{\xgm}{\mbox{$x_\gamma^{ meas}$}} 
\newcommand{\xpm}{\mbox{$x_p^{ meas}$}} 
\newcommand{\xgO}{\mbox{$x_\gamma^{\mbox{\tiny OBS}}$}} 
 
\newcommand{\ZEUSMXD}{ZEUS Collaboration, M. Derrick et al., }
\newcommand{\ltapprox}{\,\raisebox{-.4ex}{\rlap{$\sim$}}\raisebox{.4ex}{$<$}\,}
\newcommand{\bec}       {\begin{center}}
\newcommand{\eec}       {\end{center}}
\begin{document}
\thispagestyle{empty}
%
%
\begin{flushright}{\normalsize\tt DESY-97-146 \hfill ~\\July 1997 \hfill~\\} \end{flushright}
\vspace{1cm}
\bec{\Large\bf
    Observation of  Isolated High-E$_T$ Photons in Photoproduction at HERA 
     }\eec
\vspace{1cm}
\vspace{1.0cm}
\bec{\bf
    ZEUS Collaboration
    }\eec
\vspace{3cm}
%
%
\begin{abstract} \normalsize
Events containing an isolated prompt photon with high transverse energy, 
together with a balancing jet, 
have been observed for the first time in photoproduction at HERA.
The data were taken with the ZEUS detector, in a 
$\gamma p$ centre of mass energy range 120--250 GeV.  
The fraction of the incoming photon energy 
participating in the production of the prompt photon and the jet, $x_\gamma$,  shows 
a strong peak near unity, consistent with LO QCD Monte Carlo predictions.
In the transverse energy and pseudorapidity range $5\le \eTg < 10$ GeV, 
$-0.7 \le \eta^\gamma < 0.8$,  
$\eTj\ge 5$ GeV, and $-1.5\le \eta^{jet}\le 1.8$,
with $\xgO > 0.8,$ the measured cross section is 
15.3$\pm$3.8$\pm$1.8 pb, in good agreement with a
recent NLO calculation.\\[50mm]
\end{abstract}
\thispagestyle{empty}
\newpage
%
%
%
%
%
\newcommand{\address}{ }

\renewcommand{\author}{ }

\pagenumbering{Roman}

                                                                      %

\begin{center}                                                                                     
{                      \Large  The ZEUS Collaboration              }                               
\end{center}                                                                                       
  J.~Breitweg,                                                                                     
  M.~Derrick,                                                                                      
  D.~Krakauer,                                                                                     
  S.~Magill,                                                                                       
  D.~Mikunas,                                                                                      
  B.~Musgrave,                                                                                     
  J.~Repond,                                                                                       
  R.~Stanek,                                                                                       
  R.L.~Talaga,                                                                                     
  R.~Yoshida,                                                                                      
  H.~Zhang  \\                                                                                     
 {\it Argonne National Laboratory, Argonne, IL, USA}~$^{p}$
% ...
\newpage 
\vspace*{20mm}                                                                                          
% ...
$^{  23}$ now at ATLAS Collaboration, Univ. of Munich \\                                           
% ...
$^{  33}$ partially supported by the Foundation for German-Russian Collaboration                   
DFG-RFBR \\ \hspace*{3.5mm}(grant nos 436 RUS 113/248/3 and 436 RUS 113/248/2)\\                   
% ...
                                                           %
                                                           %
\newpage   
                                                           %
                                                           %
\begin{tabular}[h]{rp{14cm}}                                                                       
$^{a}$ &  supported by the Natural Sciences and Engineering Research                               
          Council of Canada (NSERC)  \\                                                            
$^{b}$ &  supported by the FCAR of Qu\'ebec, Canada  \\                                            
$^{c}$ &  supported by the German Federal Ministry for Education and                               
          Science, Research and Technology (BMBF), under contract                                  
          numbers 057BN19P, 057FR19P, 057HH19P, 057HH29P, 057SI75I \\                              
$^{d}$ &  supported by the MINERVA Gesellschaft f\"ur Forschung GmbH,                              
          the German Israeli Foundation, and the U.S.-Israel Binational                            
          Science Foundation \\                                                                    
$^{e}$ &  supported by the German Israeli Foundation, and                                          
          by the Israel Science Foundation                                                         
  \\                                                                                               
$^{f}$ &  supported by the Italian National Institute for Nuclear Physics                          
          (INFN) \\                                                                                
$^{g}$ &  supported by the Japanese Ministry of Education, Science and                             
          Culture (the Monbusho) and its grants for Scientific Research \\                         
$^{h}$ &  supported by the Korean Ministry of Education and Korea Science                          
          and Engineering Foundation  \\                                                           
$^{i}$ &  supported by the Netherlands Foundation for Research on                                  
          Matter (FOM) \\                                                                          
$^{j}$ &  supported by the Polish State Committee for Scientific                                   
          Research, grant No.~115/E-343/SPUB/P03/002/97, 2P03B10512,                               
          2P03B10612, 2P03B14212, 2P03B10412 \\                                                    
$^{k}$ &  supported by the Polish State Committee for Scientific                                   
          Research (grant No. 2P03B08308) and Foundation for                                       
          Polish-German Collaboration  \\                                                          
$^{l}$ &  partially supported by the German Federal Ministry for                                   
          Education and Science, Research and Technology (BMBF)  \\                                
$^{m}$ &  supported by the Fund for Fundamental Research of Russian Ministry                       
          for Science and Edu\-cation and by the German Federal Ministry for                       
          Education and Science, Research and Technology (BMBF) \\                                 
$^{n}$ &  supported by the Spanish Ministry of Education                                           
          and Science through funds provided by CICYT \\                                           
$^{o}$ &  supported by the Particle Physics and                                                    
          Astronomy Research Council \\                                                            
$^{p}$ &  supported by the US Department of Energy \\                                              
$^{q}$ &  supported by the US National Science Foundation \\                                       
\end{tabular}                                                                                      
                                                           %
                                                           %
\newpage

\pagestyle{plain}
\setcounter{page}{1}
\pagenumbering{arabic}
\section{Introduction} 

\begin{figure}
\centerline{\hspace*{5mm}
\epsfig{file=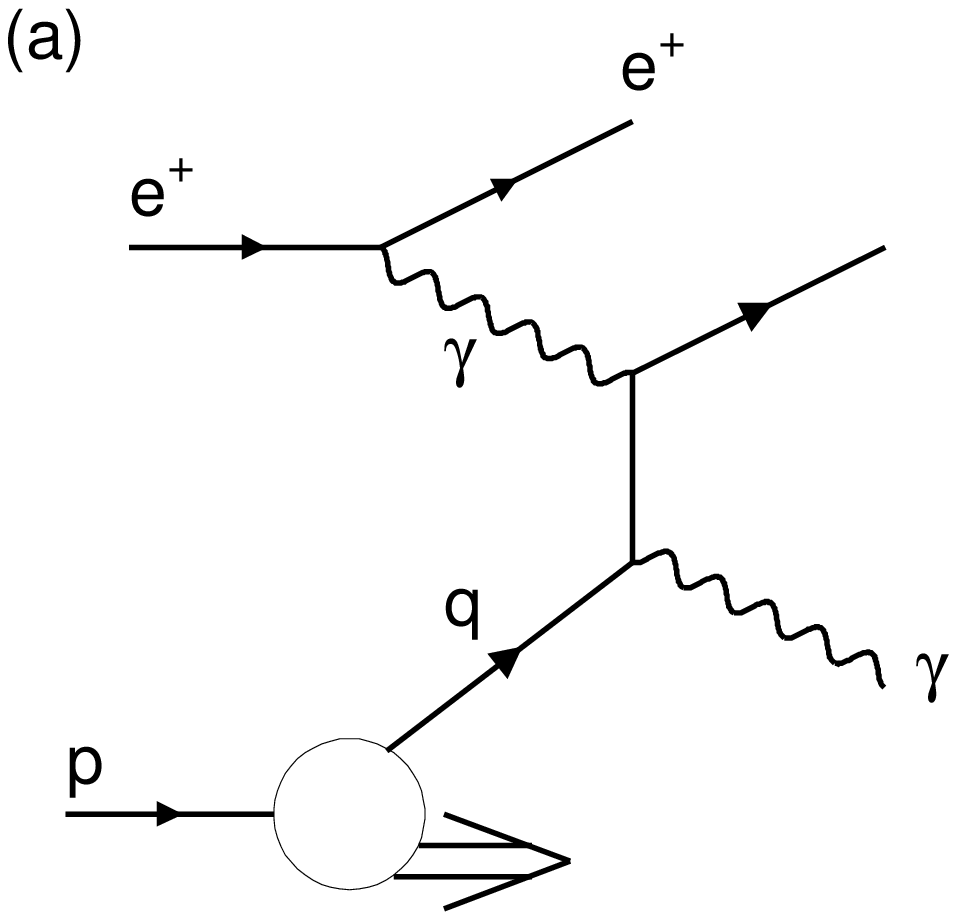,height=5.2cm,%
bbllx=141pt,bblly=200pt,bburx=483pt,bbury=488pt,clip=yes}
\epsfig{file=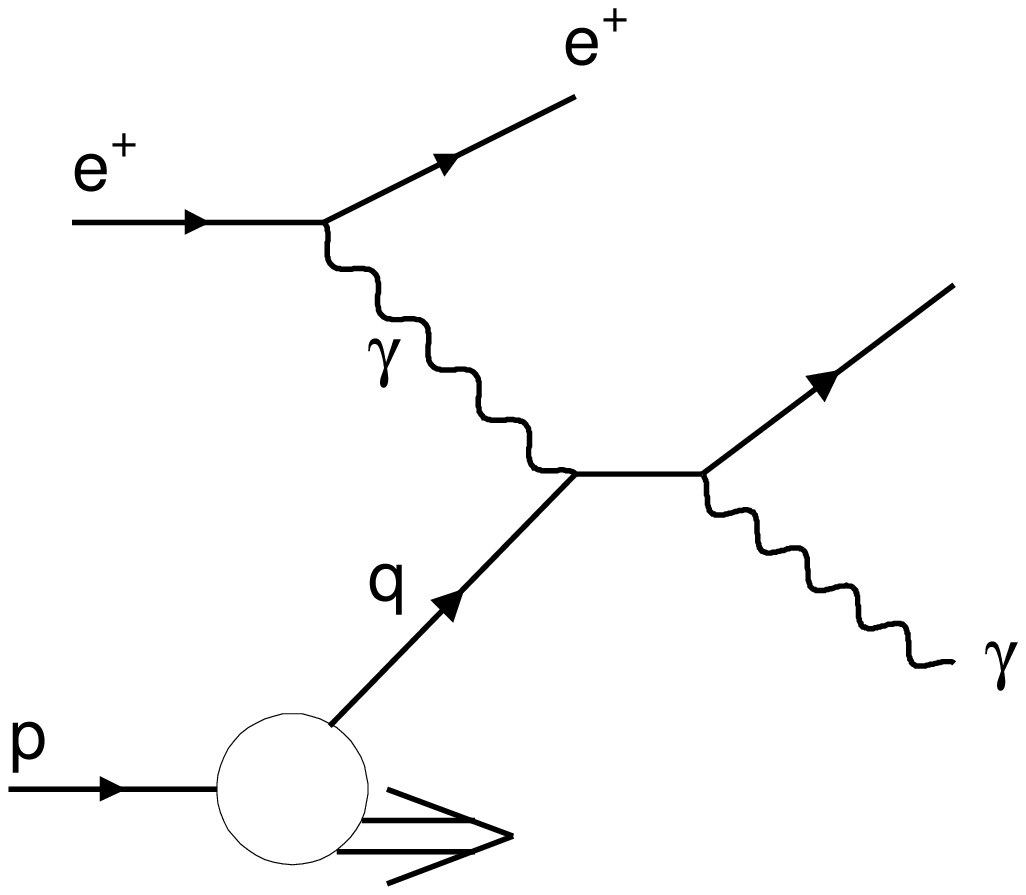,height=5.2cm,%
bbllx=141pt,bblly=200pt,bburx=483pt,bbury=488pt,clip=yes}}
\vspace{5mm}
\centerline{\hspace*{5mm}
\epsfig{file=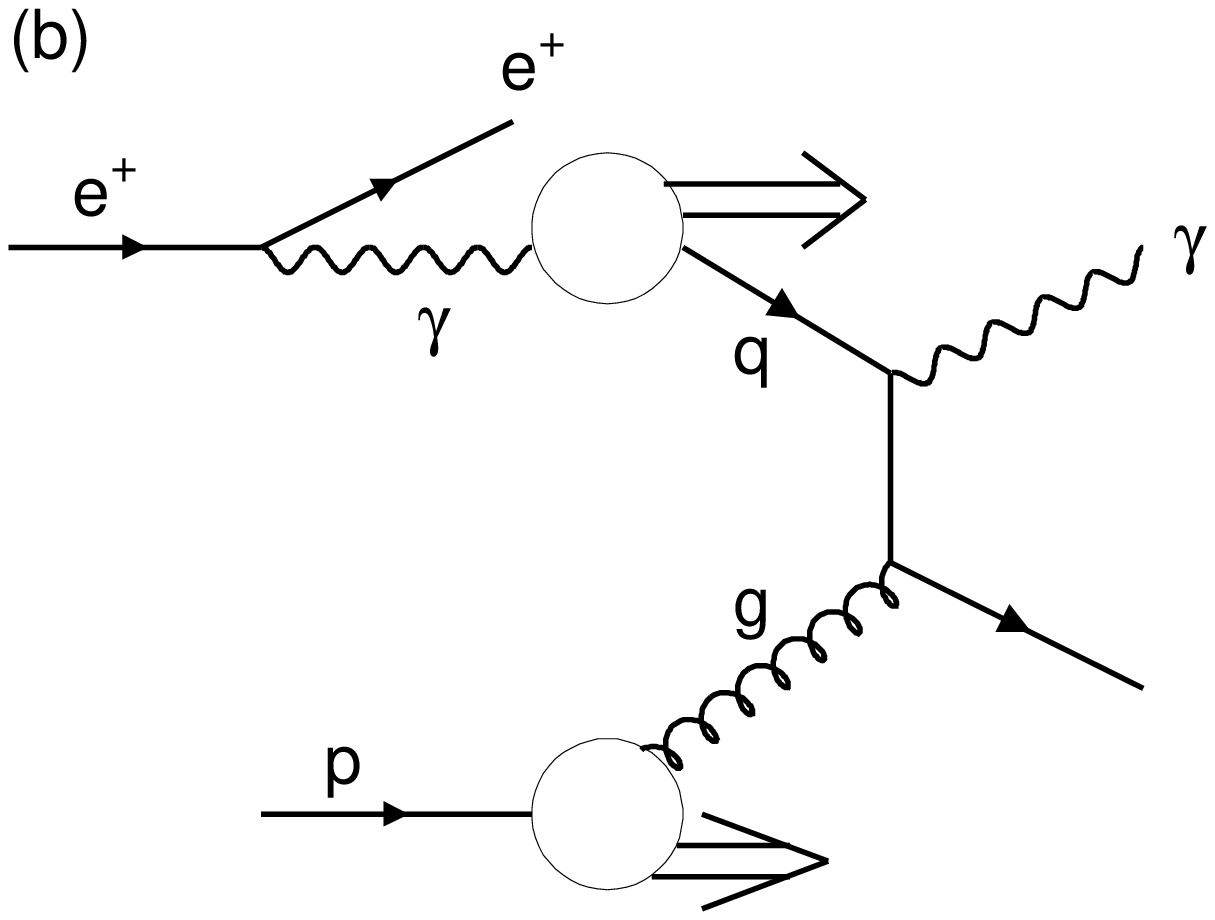,height=5.2cm,%
bbllx=24pt,bblly=200pt,bburx=414pt,bbury=488pt,clip=yes}\hspace*{10mm}
}\caption{\small  
(a) Direct LO diagrams in hard 
photoproduction producing an outgoing prompt photon.
(b) Example of resolved process.
Corresponding dijet diagrams may be 
obtained by replacing the final-state photon by a gluon.}
\label{diags}\end{figure}

A number of studies have been made at HERA on the properties of 
hard processes in quasi-real photoproduction~[1--8].
In lowest order QCD, two major types of $2\to 2$ process can be defined,
depending on how the photon interacts with a parton in the proton:
direct, in which the photon interacts 
as a pointlike particle in the hard subprocess, and 
resolved, in which the photon provides a quark or gluon which then interacts.  
The outgoing products of these subprocesses are most commonly quarks or gluons,
which at high transverse energy (\eT) can give rise to two observed jets 
(dijet events). However, final states containing a high-\eT\  jet together 
with a high-\eT\  photon are also possible (fig.~\ref{diags}).  Such 
photons are known as ``prompt"  photons 
to distinguish them from those produced via particle decays. 
In the kinematic region accessible with ZEUS, the 
direct channel in prompt photon processes is expected to be dominated by the 
direct Compton process $\gamma q \to \gamma q$, 
i.e.\ by the elastic scattering of a photon by a quark 
in the proton. The main predicted contributions to the resolved processes are 
$qg\to q\gamma$ and $q\bar q \to g\gamma$~\cite{BS}.  

A further source of prompt photons is dijet events in which an outgoing
quark radiates a high-\eT\ photon.  In measuring prompt photon processes, 
these radiative contributions are  largely suppressed
by restricting the measurement to prompt photons that are isolated from 
other particles in the event.   Such a condition is also needed in order to
reduce experimental backgrounds from neutral mesons in jets. 

In hadronic collisions, both with fixed targets~\cite{WA70etc}  and 
colliders~\cite{UAetc, Fermi}, prompt photon processes provide a 
means to study the gluon content of the proton~\cite{CTEQ,VV}.  
Fixed target studies~\cite{NA14}  have also provided a first confirmation 
of prompt photon processes in photoproduction, at a level  
consistent with QCD expectations.  At HERA, the highly asymmetric 
beam energies, together with the present detector coverage, 
restrict the sensitivity of the resolved
processes to the quark content of the photon~\cite{GS,GV,PB5} and 
the quark and gluon contents of the proton.  A particular advantage of prompt 
photon processes is that the observed final-state photon emerges from 
the hard process directly, without the intermediate hadronisation by which 
a final state quark or gluon forms an observable jet.
The cross section of the direct Compton process depends only on the
quark charge, together with the quark density in the
proton. The above considerations, together with the availability of 
next-to-leading order (NLO)  
calculations \cite{GV,Aur,Hust}, make prompt photon processes an attractive 
and relatively clean means for studying QCD, despite the low cross sections.  

From data taken in $e^+ p$ running in 1995 with the ZEUS detector at HERA, we 
have identified a class of events showing the characteristics of hard prompt 
photon processes in  quasi-real $\gamma p$ collisions.  A strong signal 
is obtained, and the presence of the direct process is clearly seen.
This is the first observation of prompt photons at $\gamma p$ centre of mass
energies an order of magnitude higher than those previously employed.
The data are compared with leading order Monte Carlo predictions.  
The cross section for the photoproduction 
of a prompt photon and a jet within a defined set of kinematic cuts is
evaluated and compared with an NLO QCD calculation.

\section{Apparatus and trigger} 

The data used in the present analysis were collected with the ZEUS detector 
at HERA. During 1995, HERA collided positrons of energy $E_e = 27.5$ GeV 
with protons of energy $E_p =820$ GeV, in 173 circulating bunches.  
Additional unpaired positron (15) and proton (7) bunches enabled monitoring of 
beam related backgrounds.  The data sample used in this analysis 
corresponds to an integrated luminosity of 6.36 pb$^{-1}$. 
The luminosity was measured by means of the positron-proton bremsstrahlung process $ep\to
e\gamma p$, using a lead-scintillator calorimeter at $Z=-107$ m\footnote{The ZEUS
coordinate system has positive-$Z$ in the proton beam direction, with
a horizontal $X$-axis pointing towards the centre of HERA. The nominal 
interaction point is at $X = Y = Z = 0.$
Pseudorapidity $\eta$ is defined as $-\ln\tan(\theta/2)$, where 
$\theta$ is the polar angle relative to the $Z$ direction. In 
the present analysis $\eta$ is always defined in the laboratory frame, and 
the $Z$ position of the event vertex is taken into account.} 
 which detects photons radiated at angles of less than 0.5 mrad to 
the positron beam direction.  

The ZEUS apparatus is described elsewhere~\cite{r1}.  Of particular importance 
in the present work are the uranium calorimeter 
and the central tracking detector (CTD). 
The calorimeter~\cite{UCAL} has an angular coverage of 99.7\% of $4\pi$ and  is divided
into three parts (FCAL, BCAL, RCAL), covering the forward (proton direction), 
central and rear polar angle ranges $2.6^{\circ}$--$36.7^{\circ}$,
$36.7^{\circ}$--$129.1^{\circ}$, and $129.1^{\circ}$--$176.2^{\circ}$, respectively. 
Each part consists of towers longitudinally subdivided
into electromagnetic (EMC) and hadronic (HAC) cells.
In test beam measurements, energy resolutions of
$\sigma_{E}/E = 0.18/\sqrt{E}$ for electrons and $\sigma_{E}/E = 0.35/\sqrt{E}$
for hadrons have been obtained (with $E$ in GeV). The calorimeter 
cells also provide time measurements which are used for beam gas background 
rejection.  The electromagnetic sections of the FCAL and RCAL comprise 
non-projective cells of transverse dimensions 20$\times$5 and 20$\times$10
cm$^2$ respectively.  
The electromagnetic section of the BCAL (BEMC) 
consists of cells of approximate dimensions $5\times20$ cm$^2$, 
with the finer dimension in the $Z$ direction.
These cells have a projective geometry so as to present a uniform 
granularity to particles emerging from the interaction point. 
In this analysis, we employ the BEMC  to identify photons with $\eT\ge 
5$ GeV. At these energies, the separation of the photons from a $\pi^0$ decay 
is of similar magnitude to the BEMC cell width,
whereas the width of a single electromagnetic shower in the BEMC is characterised
by a Moli\`ere radius of 2 cm.  The profile of electromagnetic 
signals (i.e.\ clusters of cells) in the BEMC thus gives 
a partial discrimination between those originating from single photons or 
electrons/positrons, and those originating from the decay of neutral mesons.

The CTD is a cylindrical drift chamber~\cite{CTD} situated 
inside a superconducting magnet coil which produces a 1.43 T field.
It consists of 72 cylindrical layers covering the polar angle
 region $15^{\circ} < \theta < 164^{\circ}$.
Using the tracking information from the CTD, the vertex of an event 
can be reconstructed with a
resolution of 0.4 cm in $Z$ and 0.1 cm in $X,Y$. 
In the present analysis the CTD tracks are used to locate the 
event vertex, to discriminate between high-\eT\  photons and
electrons/positrons, and in the photon isolation criterion to be described 
below.

For jet identification, a cone algorithm 
in accordance with the Snowmass Convention~\cite{DIJ3,sno} was applied to 
the calorimeter cells,  each energy deposit in a calorimeter cell being 
treated as if corresponding to a massless particle.
A cone radius $R = \sqrt{(\delta\phi)^2 + (\delta\eta)^2}$ of 1.0 was used,
where $\delta\phi, \delta\eta$ denote the distances  of the cells from the
centre of the jet in azimuth and pseudorapidity. The same algorithm 
was used online for triggering and also in the offline analysis.

The ZEUS detector uses a three-level trigger system.  The first-level
trigger used in the present analysis 
selected events on the basis of a coincidence of a regional or transverse 
energy sum in the calorimeter and at least one track in the CTD pointing
towards the interaction point.  At the second level, at least 8 GeV of
transverse energy was demanded, excluding the eight calorimeter towers
surrounding the forward beam pipe.  Cuts on calorimeter energies and timing
were imposed to suppress events arising from proton-gas collisions in the beam
pipe.  At the third level, jets were identified with $\eta^{jet}<2.5$, and at 
least two jets with $\eTj>$ 4 GeV were demanded.
These included high-\eT\  photons. 
Cosmic ray events were rejected by means of information from the tracking 
chambers and calorimeter.  An event vertex with $|Z|<60$ cm was required.  
The trigger efficiency is estimated at 97\% for the events of 
the present analysis.

\section{Event selection}
In the offline analysis, candidate prompt photon signals in the BCAL were 
selected by means of an algorithm which identified clusters of firing cells
whose energy was predominantly in the BEMC.  
The algorithm did not use tracking information, and was based on one 
developed for the identification of deep inelastic scattered (DIS) 
electrons~\cite{Ze}. Events were retained for subsequent analysis if a photon 
candidate with transverse energy $\eTg\ge 4.5$ GeV was found in the BCAL.
The BCAL requirement restricts photon candidates 
to the approximate pseudorapidity range $-0.75 < \eta^\gamma < 1.0$.  

A photon candidate was rejected if a track pointed within 0.3 radian of it; 
high-\eT\  positrons and electrons were thus removed, including the majority of
those that underwent hard radiation in the material between the interaction 
point and the BCAL.  If more than one acceptable candidate was found,
the one with highest \eT\  was taken.  Approximately 2.7k events remained
at this stage.

Events having an identified DIS positron in addition to the BCAL photon 
candidate were removed, restricting the acceptance of the pres\-ent 
analysis to incident photons of virtuality $Q^2 \ltapprox 1$ GeV$^2$.
For the remaining events, $y_{JB} = \sum(E- p_Z)/2E_e$ was calculated,  
where the sum is over all calorimeter cells, treating each signal as 
equivalent to a massless particle; i.e.\ $E$ is the energy deposited in the 
cell, and $p_Z$ is the value of $E\cos\theta$.  The quantity  
$y_{JB}$ is a measure of $y^{true} = E_{\gamma,\,in}/E_e$, where 
$E_{\gamma,\,in}$ is the energy of the incident photon.  In the case that 
an unidentified  DIS positron is present, a value of approximately unity is 
obtained.   A requirement of $0.15 < y_{JB} < 0.7$ was imposed,
the lower cut removing some residual proton-gas backgrounds and the upper
cut removing remaining DIS events.  This thereby eliminates any remaining prompt photon
candidates which were in actual fact misidentified DIS positrons.  Wide-angle 
QED Compton scatters ($e(p)\to e\gamma(p)$) were also excluded by this cut.

A jet with transverse energy $\eTj>$ 4.5 GeV and pseudorapidity 
$-1.5\le \eta^{jet}\le 1.8$ was also demanded.  If more than one
such jet was found, that with the highest transverse energy  
was used in the analysis.

An isolation cone was now imposed around the photon candidate:
within a cone of unit radius in $(\eta,\,\phi)$, the total \eT\  from 
other particles was required not to exceed  $0.1\eTg $. 
This was calculated by summing the \eT\  in each calorimeter cell within the 
isolation cone, treating each cell energy as equivalent to that of
a massless particle. Additional contributions were included from charged tracks 
which started within the isolation cone but curved out of it; the small 
number of tracks which curved into the isolation cone were ignored. 
This isolation condition greatly reduces the dijet background, by removing the 
large majority ($\approx$80\%) of events where the photon candidate 
is associated with a jet, and is therefore either hadronic in 
origin or else a photon radiated within a jet.
In particular, as discussed by previous
authors~\cite{GV}, it removes most dijet events in which a 
photon is radiated from a final state quark.
The losses of direct and resolved prompt photon events 
due to the isolation condition were 
found from Monte Carlo studies to be $\approx$5\%
and $\approx$17\% respectively.  Overall, the isolation condition 
removed 70\% of the candidates remaining at this stage, leaving 568 events.

Some tighter kinematic conditions were finally applied.
The photon candidate was required to have $5 \le \eTg < 10$ GeV, the 
upper limit being imposed due to the increasing difficulties in distinguishing 
photons from $\pi^0$ in the BEMC above this energy.  As will be seen below, 
the photons and jets were found to be azimuthally back-to-back in the 
detector, and no evidence for a photon signal was found 
in the candidates where the azimuthal separation $\Delta\phi$ 
between the photon candidate and the jet was less than $140^\circ$. 
$\Delta\phi$ was therefore required to be above 140$^\circ$ for the final
event sample. 
The number of events remaining at this stage was 256. 

\section{Analysis method} 

\subsection{Monte Carlo simulations}

Three types of Monte Carlo samples were employed in this analysis
to simulate: (1) the
prompt photon processes, (2) single particles ($\gamma$, $\pi^0$, $\eta$), 
and (3) dijet processes that could mimic a prompt photon final state. All 
generated events were passed through a full simulation of the ZEUS detector.

The PYTHIA-5.7~\cite{Pythia} Monte Carlo generator 
was used to simulate the direct and
resolved prompt photon processes and dijet processes. The 
MRSA proton structure function and GRV(LO) photon structure function 
were employed. The minimum $p_T$  of the hard scatter was set to 2.5 GeV and 
the maximum $Q^2$ set to 4 GeV$^2$. In running PYTHIA, initial and final
state QCD and QED radiation were turned on.  
Multiple interactions were not included in the resolved event sample 
since they are not expected to have a significant effect in the present
analysis.

Three additional Monte Carlo data sets were generated, comprising 
large samples of single $\gamma$, $\pi^0$  and $\eta$ respectively. The single particles were 
generated over the acceptance of the BCAL with a flat transverse energy distribution 
between 3 and 20 GeV; \eT-dependent exponential weighting functions  
were subsequently applied to reproduce the observed \eT\  distributions. 
These samples are important for the understanding and the separation of 
signal and background using shower shapes in the calorimeter. 

To produce a Monte Carlo sample for background studies, direct and resolved 
dijet events were generated.  Event samples of this kind also enabled the 
radiative contribution to the prompt photon signal to be evaluated.
Measurements from ALEPH~\cite{Aleph}  have shown that the 
PYTHIA generator gives a qualitative description of this 
type of process in $e^+e^-$ annihilation,
but that the magnitude may be underestimated.

\subsection{Identification of photon signal} 

Electromagnetic signals in the calorimeter
that do not arise from charged particles arise predominantly from photons 
and from $\pi^0$ and $\eta$ mesons.  The minimum distance at the BCAL  
between the photons from a $\pi^0$ with $\eT = 5$ (10) GeV is 
7 (3.5) cm, which is comparable to the $Z$ width of the BEMC cells
and smaller than the azimuthal cell width. 
Thus it is not normally possible to resolve the two 
photons from a $\pi^0$ and hence reconstruct its decay.  The $\eta\to 2\gamma$
decay angle, although broader by a factor of three than the $\pi^0$, 
is still unresolvable in most cases, while the $\eta\to3\pi^0$ mode gives 
up to six photon signals which may be varyingly merged together.
It is therefore not possible to distinguish single photons from the 
$\pi^0$ and  $\eta$ backgrounds on an event-by-event basis.

A typical high-\eT\  photon candidate in the BEMC consists  of a cluster of
4-5 cells selected by the electron finder.
Two shape-dependent quantities were studied in order to distinguish
photon, $\pi^0$ and $\eta$ signals.  These were (i)
the mean width \mean{\delta Z}\ of the BEMC cluster in $Z$ and (ii) the fraction
$f_{max}$ of the photon candidate energy found in the most energetic
cell in the cluster. \mean{\delta Z}\ is defined as the mean 
absolute deviation in $Z$ of the cells in the cluster, energy weighted, 
measured from the energy weighted mean $Z$ value of the cells in 
the cluster. It is expressed in units of the BEMC cell width in the 
$Z$ direction. 
  
\begin{figure}\centerline{
\epsfig{file=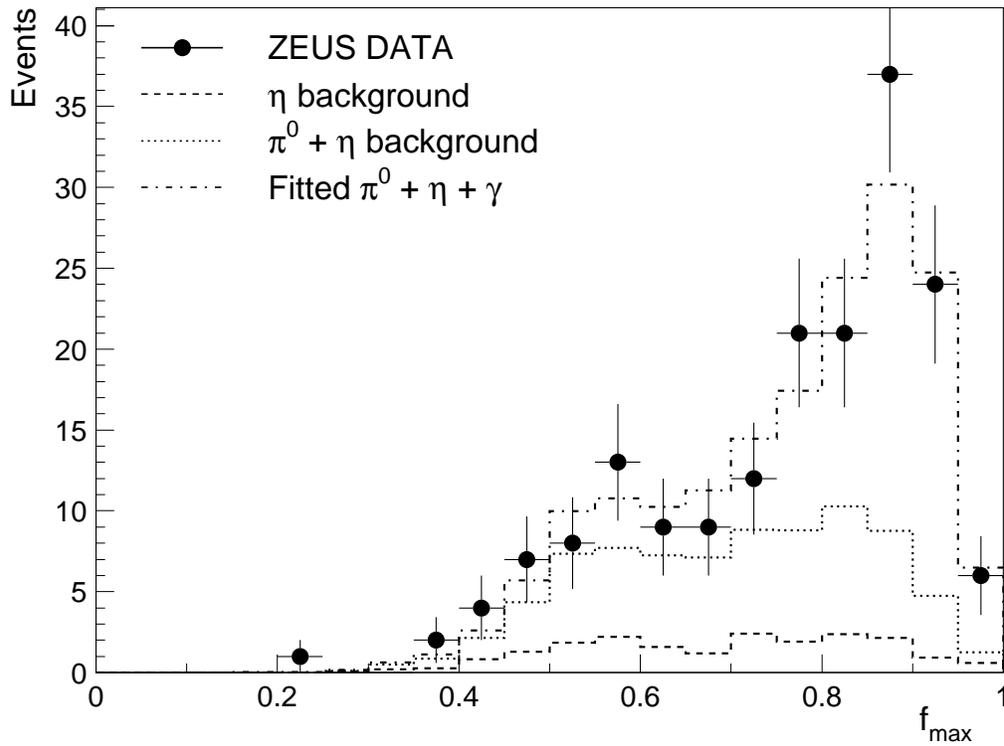,%
height=100mm,bbllx=17pt,bblly=096pt,bburx=530pt,bbury=450pt,clip=} }
\caption{\small Distribution of $f_{max}$            
for prompt photon candidates in selected events, including a requirement that 
$ \protect\mean{\delta Z}\, < \, 0.65 $ cell widths.
Also plotted are fitted Monte Carlo curves for photons, $\pi^0$ and $\eta$ 
mesons with similar selection cuts as for the observed photon candidates
(see text).}
\label{fmax}\end{figure}

From the Monte Carlo samples of single $\gamma$, $\pi^0$ and $\eta$ in the 
BCAL, it was established that the photon and $\pi^0$ signals both had small  
probabilities of having $ \mean{\delta Z}  \ge 0.65$.  A cut was therefore 
imposed at this value, separating candidates with $ \mean{\delta Z}  \ge 0.65$, 
taken to be mainly $\eta$ mesons, from those in the lower 
\mean{\delta Z}\  range, which 
comprised mainly photons and $\pi^0$ mesons with a small admixture of $\eta$.  
The  $f_{max}$ distribution for the sample of events with 
$ \mean{\delta Z} < 0.65$ is shown in fig.~\ref{fmax}.  A fit to a mixture 
of  $\gamma$, $\pi^0$ and $\eta$ was performed on this $f_{max}$ distribution, together 
with the numbers of events 
with $ \mean{\delta Z} \ge 0.65$, which determined the $\eta$ contribution. 
From the fit it is evident that the $\eta$ and 
$\pi^0$ $f_{max}$ distributions are similar in shape, whereas the photon 
$f_{max}$ distribution has a sharp peak above a value 0.75.  The 
fit to the experimental $f_{max}$ distribution is good, and 
above 0.75 the data are dominated by a substantial photon component. 

As a check on the procedure, the $f_{max}$ distribution of deep inelastic
scattered positrons in the BCAL was compared with a corresponding Monte Carlo 
sample.  A small discrepancy in the mean positions of the 
two peaks was found; as a consequence all the plotted 
Monte Carlo $f_{max}$ values have been scaled by a factor 1.025$\pm$0.005.
This gives rise to a systematic uncertainty in the final results of $\pm$4\%.

We perform a background subtraction on the assumption that the data may be
expressed as a sum of photon signal plus neutral meson background
as indicated in fig.~\ref{fmax}.
An important conclusion from fig.~\ref{fmax} is that the shape of the $f_{max}$ 
distribution is similar for the $\eta$ and  $\pi^0$ contributions.
It follows that the background subtraction is 
insensitive to uncertainties in the ratio of $\pi^0$ to $\eta$ in the fit.

\subsection{Signal/background separation}

Guided by fig.~\ref{fmax}, we divide the data into two subsamples, 
consisting of events whose photon candidate has 
$f_{max} \ge 0.75$ and $f_{max} < 0.75$ respectively. 
Such subsamples are respectively enriched and impoverished
in events containing a genuine high-\eT\  photon, and will be referred
to as ``good" and ``poor" subsamples.  In any given bin of a physical
quantity of interest, using the same
definition, the events can be likewise divided into good and poor subsamples. 
 Let these consist of $n_{good}$ and $n_{poor}$ events respectively.
The  values of $n_{good}$ and $n_{poor}$ in a bin 
may be written:
\newlength{\mm}\newlength{\mmm}\newlength{\bb}\settowidth{\mm}{$\alpha$}
\settowidth{\mmm}{$(1 - \alpha$\hspace*{-\mm}}\settowidth{\bb}{)}
\begin{eqnarray}
n_{good} &=& \hspace*{0.5\mmm}\alpha\hspace*{\bb}n_{sig}\hspace*{0.5\mmm} + 
\hspace*{0.5\mmm}\beta\hspace*{\bb}n_{bgd}\hspace*{0.5\mmm} \nonumber \\
n_{poor} &=& (1 - \alpha) n_{sig} + (1-\beta) n_{bgd} \end{eqnarray}
where $n_{sig}, \:n_{bgd}$ are numbers of signal
(i.e.\ photon) and background (i.e.\ $\pi^0$ or $\eta$) events in
the bin.  The coefficients $\alpha$ $(\beta)$ are the probabilities
that a signal (background) event will end up in the good subsample.
They are evaluated from the known shapes of the Monte Carlo $f_{max}$ 
distributions of the photons and of the fitted $\pi^0 + \eta$ background, 
as shown in fig.~\ref{fmax}.  For given observed values of
$n_{good}$ and $n_{poor}$  it is now straightforward to solve (1) for
the values of $n_{sig}$ and $n_{bgd}$, and to evaluate their errors.  

\section{Results} 
\begin{figure}\centerline{ \epsfig{file=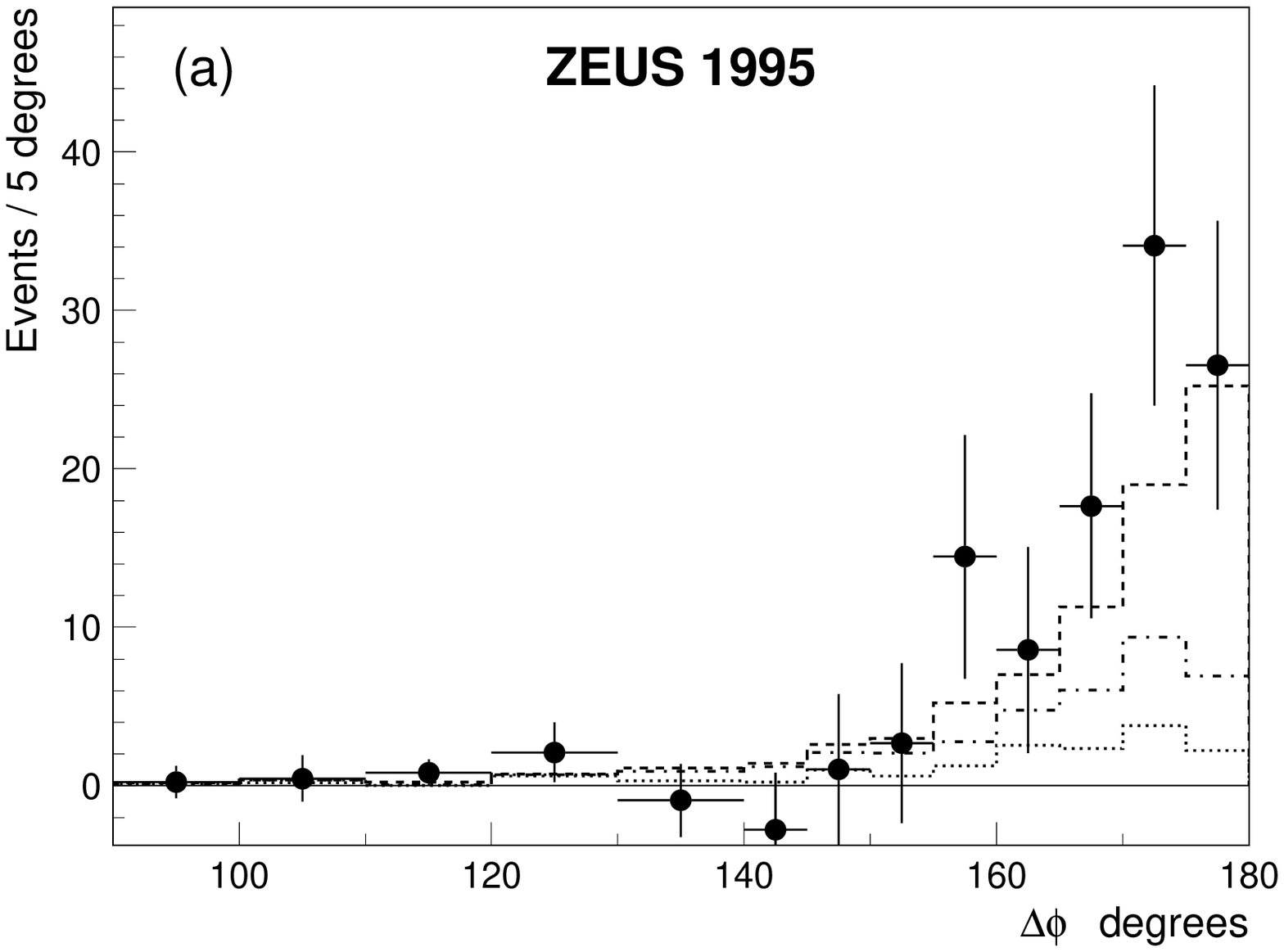,%
height=092mm,bbllx=70pt,bblly=070pt,bburx=550pt,bbury=430pt,clip=} }
\vspace*{5mm}
\centerline{ \epsfig{file=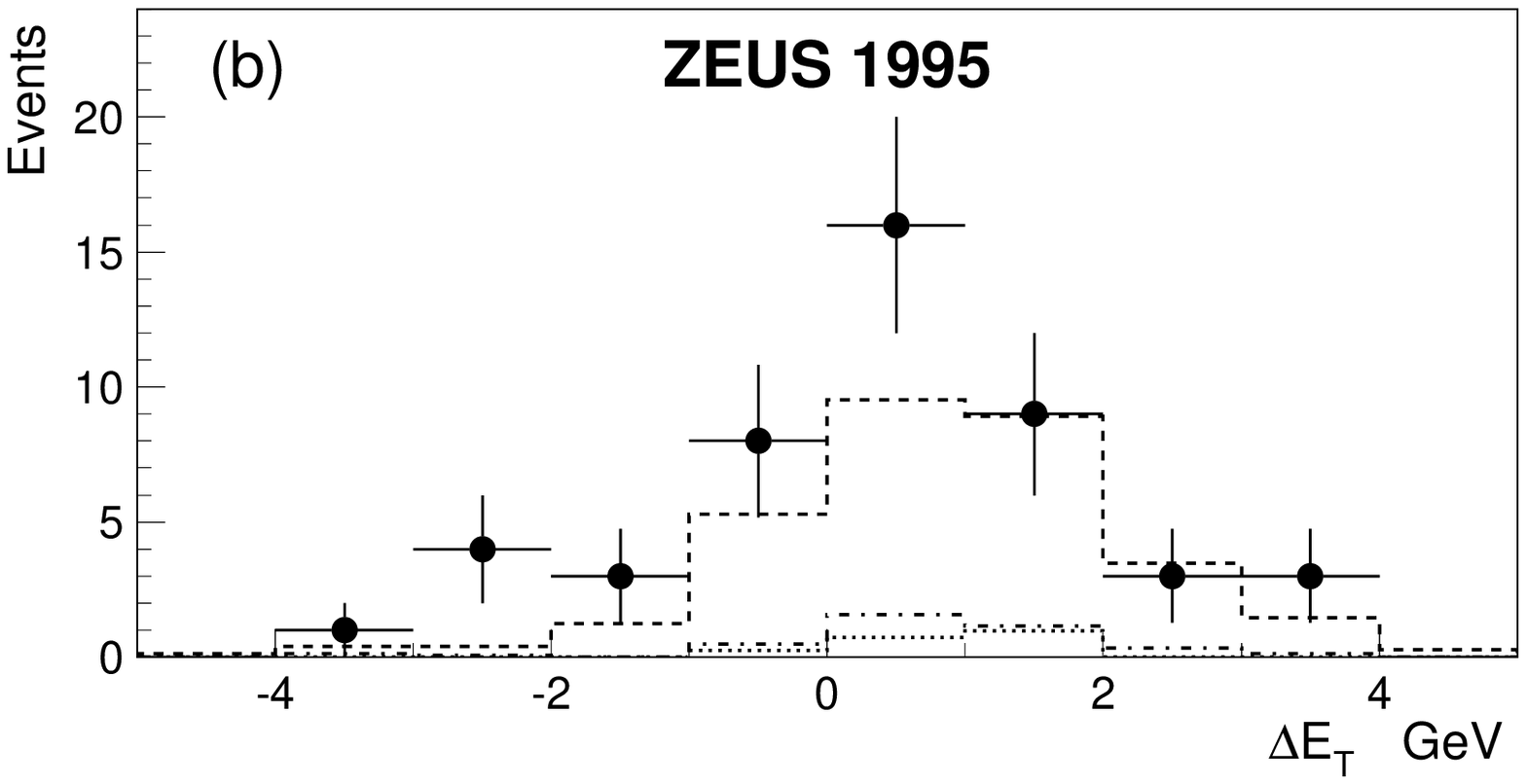,%
height=064mm,bbllx=70pt,bblly=70pt,bburx=550pt,bbury=320pt,clip=} }
\caption{\small 
(a) Background subtracted distribution in $\Delta\phi$ 
for photon-jet pairs, before application of cut on $\Delta\phi$. 
(b) Distribution in $\Delta \eT = (\eTg -  \eTj)$ 
for selected events ($\xgm > 0.9$, c.f.\ fig.~\protect\ref{results}). 
Points = data; dotted histogram = MC radiative contribution; dash-dotted = 
radiative  + resolved; dashed = radiative + resolved + direct.}
\label{det}\end{figure}

\begin{figure}
\vspace{-14mm}
\centerline{
\epsfig{file=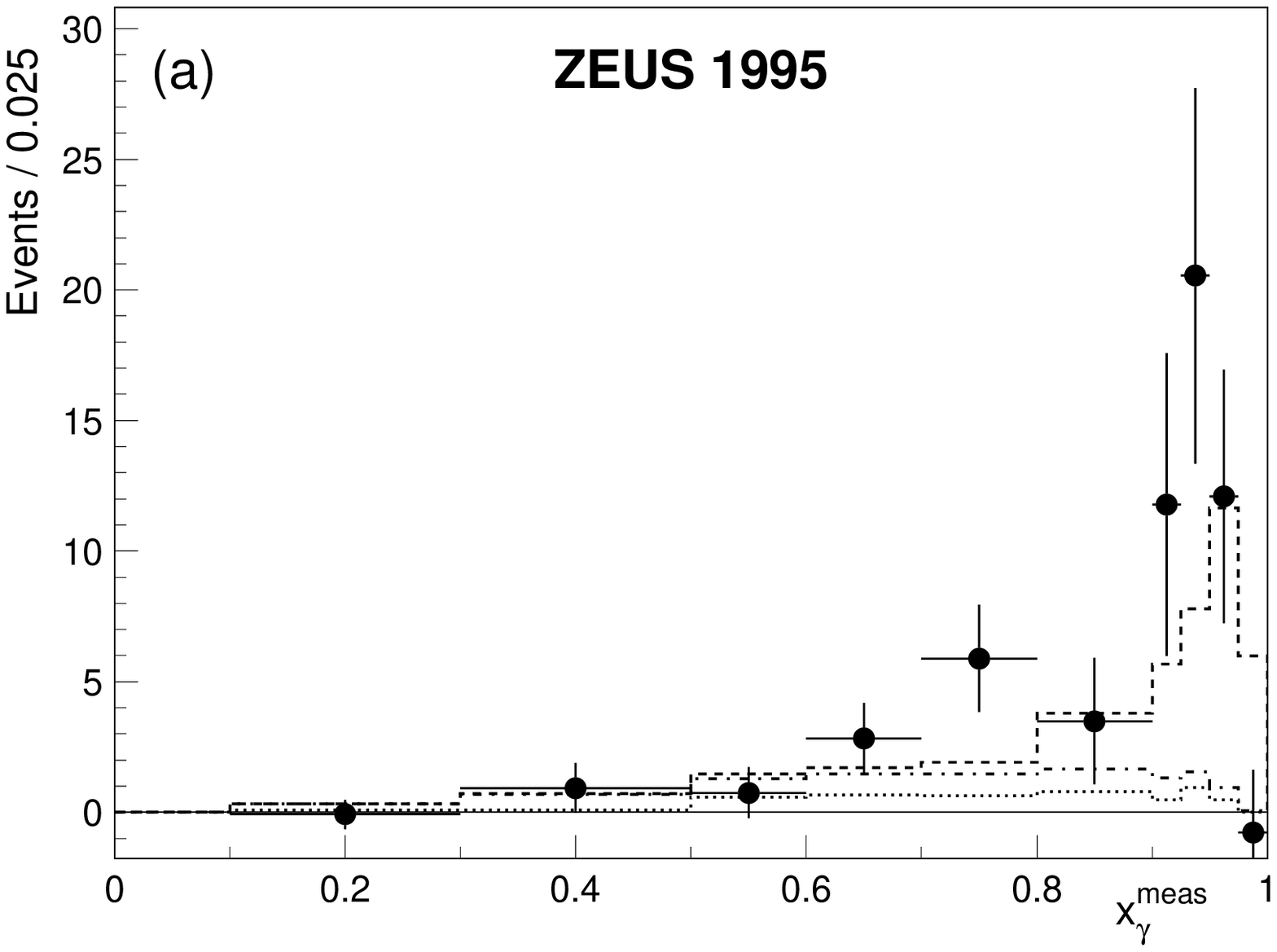,height=9.2cm,%
bbllx=70pt,bblly=66pt,bburx=537pt,bbury=427pt,clip=yes} }
\vspace{-5mm}
\centerline{
\epsfig{file=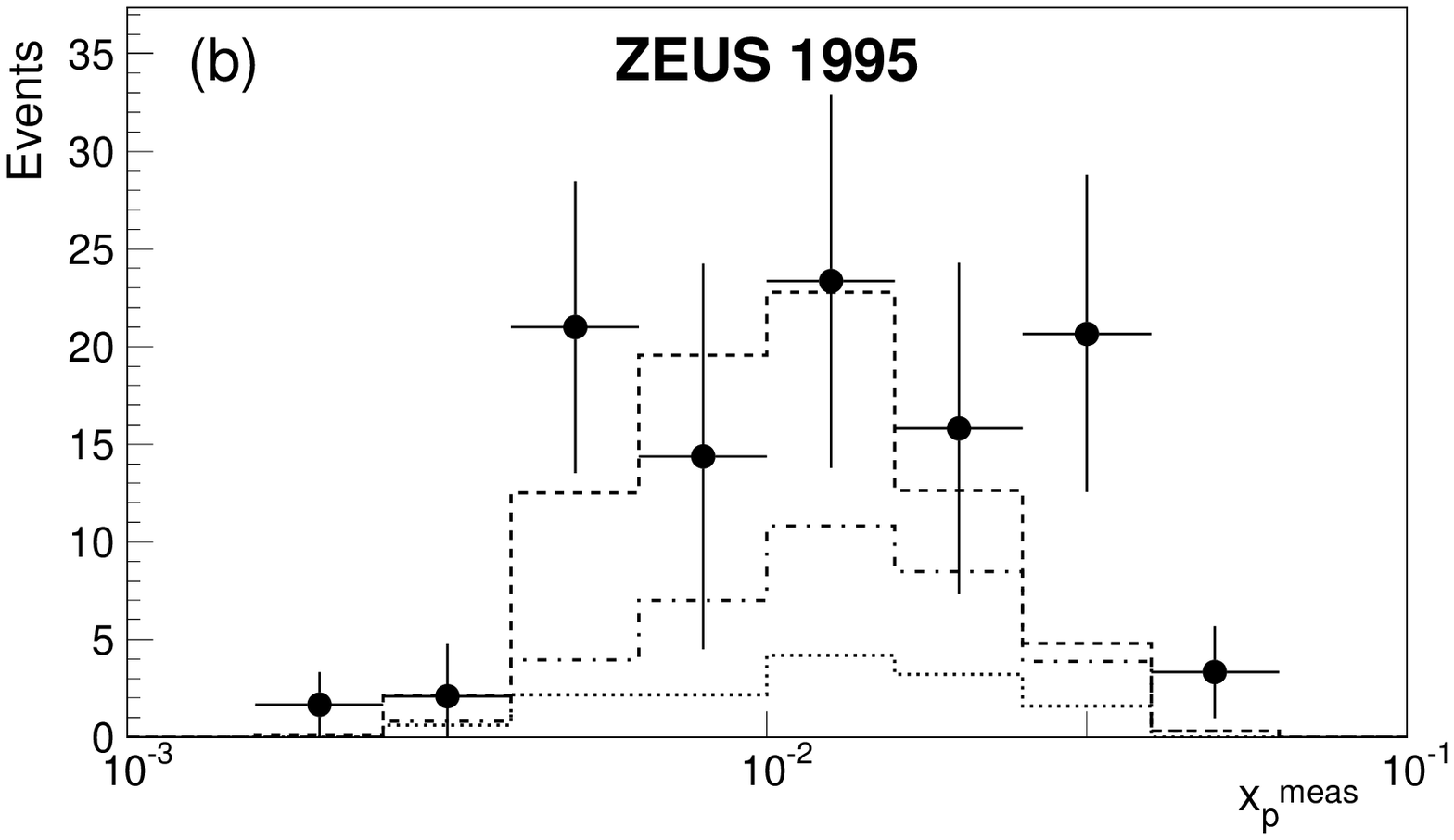,height=7.0cm,%
bbllx=70pt,bblly=96pt,bburx=537pt,bbury=370pt,clip=yes} }
\vspace{5mm}
\caption{\small (a) Distribution in \xgm\ of prompt photon events after 
background subtraction. 
Points = data; dotted histogram = MC radiative contribution; 
dash-dotted = radiative 
+ resolved; dashed = radiative + resolved + direct.
Plotted values represent numbers of events per 0.025    
interval of \xgm; i.e.\ 
total number of events in bin = plotted value $\times$ bin width $/$ 0.025.
Errors are statistical only and no corrections have been applied to the data. 
(b) Distribution in \xpm, data and histograms as (a); 
the plotted points are events per bin.  
}\label{results} 
\end{figure}

Fig.\ \ref{det}(a) shows that the background-subtracted distribution 
of the azimuthal angle difference ($\Delta\phi$) between the photon and the accompanying jet 
is well peaked at 180$^\circ$ as expected.  
Also shown are the corresponding Monte Carlo expectations for contributions from: 
(i) dijet events in which an outgoing quark radiates an isolated 
high-\eT\  photon; 
(ii) the resolved prompt photon processes; 
(iii) direct prompt photon production.
Here and in fig.~\ref{results}, the Monte Carlo distributions are 
normalised to the same integrated luminosity as the data.
For clarity, the plotted histograms show (i), (i)+(ii) and (i)+(ii)+(iii). 
The error bars on the data points are statistical only.  Reasonable agreement
between the data and the sum of the Monte Carlo distributions  
is seen. Fig.\ \ref{det}(b) shows the difference in transverse energy between 
the photon and the jet for the virtually background-free class of  
events with $\xgm > 0.9$ (see below).
The \eT\ of the photon and jet are on average well balanced,
and the shape of the Monte Carlo satisfactorily reproduces that of the data.

The fraction $x_\gamma$ of the incoming photon momentum 
that contributes to the production of the high-\eT\ photon and jet is 
studied. For the case of a direct process (as defined at leading order), 
$x_\gamma = 1$.  Here we estimate the ``observable" quantity 
$\xgO= (E_{T\,1}e^{-\eta_1}+E_{T\,2}e^{-\eta_2})/2E_ey^{true}$~\cite{DIJ3} 
as 
$$\xgm=\sum_{\gamma,\, jet}(E-p_Z)\left/ 2E_e\,y_{JB}.\rule{0ex}{2ex}\right.$$
The sums are over the photon candidate and the 
calorimeter cells in the jet, each signal being treated as equivalent to 
that of a massless particle of energy $E$ and longitudinal momentum component 
$p_Z$.  The distribution of the resulting signal is shown in 
fig.~\ref{results}(a).  A narrow peak is seen 
in the signal near $\xgm=1,$ which we identify with the direct Compton process.  
The bin widths at the peak are chosen to be comparable to the 
measurement resolution on \xgm\ in this region (0.035), 
and are governed by the statistics elsewhere.
The Monte Carlo distribution is similar in shape and magnitude.  
QCD radiation, hadronisation outside the jet cone and detector effects  
lower the peak position slightly from its expected value of unity.  There is, 
in addition, a tail of entries extending over lower \xgm\ values. 
It is not possible to draw conclusions concerning the presence of a resolved 
photon component, except to remark that the observed numbers of
events at low $\xgm$ are consistent with the level expected from the Monte
Carlo.  The predicted radiative contribution is not negligible compared to 
the resolved contribution.  
For $\xgm>0.8$, the number of events in the signal is 57.6;
the size of the signal is insensitive to small
variations in the isolation conditions.
The Monte Carlo calculations indicate that approximately 75\% of the events  
in this region are direct Compton, 12\% are resolved and 13\% are radiative.
It should be noted that the effects of higher order 
QCD processes are not included in the present Monte Carlo simulation.  

The background distribution is consistent with zero for $\xgm > 0.9$. 
Below this value it averages at two counts per 0.025 interval of \xgm.
As a test of the 
procedure, an identical analysis was performed using the 
photon candidates obtained in the Monte Carlo dijet samples, excluding 
the events where an outgoing quark radiates a photon. These candidates were 
due to a physically realistic mixture of simulated neutral mesons.
Here, results were obtained consistent with zero prompt photon signal,
accompanied by the expected finite background.

The simulations make use of parton densities in the proton 
in a region of $x_p$ (the fraction of the proton's momentum 
entering the hard process) where these are experimentally well known.
A value of $x_p$ is calculated analogously to \xgm\ as
$\xpm=\sum_{\gamma,\, jet}(E+p_Z)/2E_p$. 
Fig.~\ref{results}(b) shows the distribution of \xpm\  
compared to the predictions from PYTHIA.  Again, reasonable agreement 
between experiment and Monte Carlo is seen. 

A systematic uncertainty exists in the
comparison of the data with the Monte Carlo, due to an estimated 3\%
uncertainty on the calorimeter calibration. 
Rescaling the calorimeter cell energies in the direct prompt photon MC by 
$\pm$3\% changes the number of events accepted by $\pm$8\%. The \eT\ weighting 
of the generated single particles was also varied by amounts corresponding 
to the experimental uncertainty on the \eT\  distributions of these particles.
In each case, the results were affected by approximately 1\%. The $f_{max}$ distributions 
exhibited by the single particle MC samples, and by the data, varied little with the
pseudorapidity of the particles.  To test for sensitivity to this, an analysis was
performed using Monte Carlo photons generated 
only in the range $|\eta^\gamma|>0.35$, and 
applying the $f_{max}$ distribution of this sample to the data in 
place of the standard distribution.  This changed the results by $\approx$3\%.   
It is possible that the $\eta : \pi^0$ ratio in the background might vary 
from bin to bin in a given physical quantity. To evaluate the possible
effects of this, we performed a fit in which the 
$\eta : \pi^0$ ratio in the background was halved from its 
normal value.  This altered the final results by 
about 1\%.  

A cross section for the prompt photon process is  evaluated as follows.
The number $n_0$ of observed events in a given \xgm\ range is determined,
subject to the experimental selection conditions stated above.  From 
Monte Carlo event samples, we evaluate the number of fully 
reconstructed events subject to the same conditions as the data ($n_1$), 
and the number of events satisfying a set of defined conditions at the 
final-state hadron level ($n_2$).   These latter conditions are chosen, 
taking experimental effects into account, to be approximately equivalent 
to the selection conditions on the data. 
The ratio $n_2/n_1$ then represents a correction factor to be 
applied to $n_0$, to give an experimental cross section for the process 
defined by the conditions used to evaluate $n_2$.

We quote a cross section for the process 
$$e p \to e + \gamma_{prompt} + \mbox{\it jet} + X$$ 
with $\xgO\ge 0.8$.  Here, $\gamma_{prompt}$ denotes 
a final state isolated prompt photon with $5\le \eTg < 10$ GeV 
and $-0.7 \le \eta^\gamma < 0.8$ ; {\it jet\/} denotes a 
``hadron jet" with $\eTj\ge 5$ GeV and $-1.5\le \eta^{jet}\le 1.8$.
$X$ includes the proton remnant, a possible photon remnant
and any other final state products. 
A ``hadron jet" is a jet constructed out of the 
primary final state particles (charged or uncharged) in a given Monte Carlo  
generated  event.  The energy and direction of each primary final state 
particle are used in the jet finder in the same way as the calorimeter cells 
are used in the experimental jet finding.  
Limits of $0.16 < y^{true} < 0.8$ and $Q^2 < 1$ GeV$^2$ were applied 
in the hadron level event definition, and an equivalent isolation cone 
definition was applied as in the data, i.e.\ the total \eT\ from 
other particles inside a cone of unit radius around the generated prompt photon was
permitted to be at most 0.1 of that of the photon.
The systematic errors are dominated by the 8\% contribution due to the
uncertainty in the calorimeter calibration.   The correction factor $(n_2/n_1)$
is $1.69\pm0.09$, where the error comes from Monte Carlo statistics and the 
uncertainty in the size of the radiative contribution.
A contribution of 5\% is included to allow for differences 
between the shapes of the data and Monte Carlo distributions. 
As a systematic check on the Monte Carlo simulation of the noise in the
calorimeter cells due to uranium radioactivity, the minimum energy deposit in a 
cell above which the cell enters the analysis was varied.  The change in the 
cross section was 2\%.

With the above definitions, a cross section of 15.3$\pm$3.8$\pm$1.8 pb
is obtained, where the errors are statistical and systematic respectively.
This result can be compared with  NLO calculations at the parton level 
of Gordon~\cite{Gor} (using an LO radiative contribution),  
in which the integrated cross section for the process 
$e p \to e + \gamma_{prompt} + \mbox{\it jet} + X$ 
is evaluated under the same kinematic conditions as used above. Using 
the GS and GRV NLO~\cite{GRV} photon parton densities, integrated cross sections
of 14.05 (13.17) pb and  17.93 (16.58) pb respectively are obtained,
where the first value is calculated at a QCD scale $\mu = 0.25\,(\eTg)^2$
and the second, parenthesized value at $\mu = (\eTg)^2$.
These values cover the range of theoretical uncertainty of each calculation.
The experiment and theory are in good agreement. 
The ratio of the resolved to the direct contribution in 
the calculated cross section is scale dependent, and takes values in the range 
0.23--0.34 for the GS parton densities. 

\section{Conclusions}
We have observed for the first time isolated high-\eT\ photons, 
accompanied by balancing jets, in photoproduction at HERA.
The \xgm\ distribution of the events is in good general agreement with 
LO QCD expectations as calculated using PYTHIA.
In particular, a pronounced peak at high \xgm\ is observed, 
indicating the presence of a direct process. 

We have measured the cross section for prompt photon production 
in $ep$ collisions satisfying the conditions of having (i) an 
isolated final-state photon with  $5\le \eTg < 10$ GeV, 
accompanied by a jet with $\eTj\ge 5$ GeV, (ii) the photon and jet lying 
within the respective 
laboratory pseudorapidity ranges (--0.7, 0.8) and (--1.5, 1.8),
(iii) $\xgO \ge 0.8$, (iv) $0.16 < y^{true}< 0.8$, (v) $Q^2 < 1$ GeV$^2$.
The value obtained is  15.3$\pm$3.8$\pm$1.8 pb, in good agreement with a
recent NLO calculation of the process.  

\vspace{8mm}

\noindent
{\Large \bf Acknowledgements.}\\[3mm]
It a pleasure to thank the DESY directorate and staff for their
support and encouragement.  The outstanding efforts of the HERA
machine group in providing improved luminosities in 1995 are 
much appreciated.  We are extremely grateful to L. E. Gordon for 
helpful conversations, and for making available
to us calculations of the NLO prompt photon cross section.

\newpage

\newpage
\pagestyle{empty}

\end{document}